\documentclass[arxiv, usenatbib]{iupartex}

\usepackage{newtxtext,newtxmath}
\usepackage[T1]{fontenc}

\title[Colour-Magnitude Relations]{Investigating the Colour-Magnitude Relations for White Dwarf Stars in SDSS Photometry}

\author[Caliskan and Gokmen]{%
O. Caliskan$^{1\cc}$,\orcid{0009-0003-5839-8007}
and
S. Gokmen\orcid{0000-0002-0108-4782}
\affsep \\
$^1$Programme of Astronomy and Space Sciences, Institute of Graduate Studies in Science, Istanbul University,  34116, Istanbul, T\"{u}rkiye\\
$^2$ Department of Physics, Florida Atlantic University, Boca Raton, FL 33431, USA
}

\corres{O. Caliskan}{caliskanozcan@hotmail.com, ozcan.caliskan@ogr.iu.edu.tr}

\pubyear{2025}

\doiheader{XXXXXXX/PAR.20XX.00000}
\date{
	\pSubmit{00.00.0000} 
	\pRevReq{00.00.0000}
	\pLastRevRec{00.00.0000}
	\pAccept{00.00.0000}
	\pPubOnl{00.00.0000}
}
\volume{0}
\volnumber{1}

\begin{document}
\label{firstpage}
\pagerange{\pageref*{firstpage}--\pageref*{lastpage}}
\maketitle

\begin{abstract}

In this study, colour-magnitude relations (CMRs) for DA-type white dwarfs in the Sloan Digital Sky Survey (SDSS) photometric system were investigated. For this purpose, the SDSS data for 20,247 white dwarf stars, as provided in the study by \citet{Anguiano17}, were matched with the {\it Gaia} third data release ({\it Gaia} DR3) catalogue to obtain trigonometric parallax ($\varpi$) data. The SDSS {\it ugriz} magnitudes of the stars were corrected for interstellar extinction using dust maps provided for the Milky Way, and distances from the Sun to the stars were calculated. The SDSS magnitudes were thus corrected for the effects of interstellar extinction. For the calibration of the stars, 5,516 white dwarf stars were selected, with apparent magnitudes brighter than $g_0=21$ mag and relative parallax errors measured to better than $\sigma_\varpi/\varpi=0.1$. Subsequently, three separate CMRs were derived for the absolute magnitudes $M_{\rm g}$, $M_{\rm r}$, and $M_{\rm i}$, each calibrated to two-colour indices. The coefficient of determination ($R^2$) of the obtained CMRs are highly reliable in the bf range of 0.86 to 0.95. Moreover, the standard deviations of the differences between the absolute magnitudes obtained from the relations and the original ones of the calibration stars range from 0.26 to 0.37 mag.
\end{abstract}




\section{Introduction}

White dwarf stars are remnants of low- to intermediate-mass stars that have exhausted their nuclear fuel and undergone a cooling and contraction process. They play a crucial role in stellar evolution, providing valuable insights into the late stages of stellar life cycles. Over the years, numerous studies have focused on characterising the properties of white dwarfs, including their mass, radius, temperature, and luminosity. Early studies, such as that of \citet{Koester79} and \citet{Shipman79}, established the fundamental mass-luminosity relationship for white dwarfs, which has since been refined with more precise observational data. Advances in photometry, particularly through surveys such as the Sloan Digital Sky Survey \citep[SDSS,][]{York20} and the {\it Gaia} mission \citep{Gaia16mission}, have significantly improved the determination of the parameters of white dwarfs. \citet{Kleinman04} used SDSS data to derive a comprehensive catalogue of white dwarfs, allowing the identification of their cooling ages and spectroscopic characteristics. Additionally, thanks to the different data releases provided by the {\it Gaia} satellite, it has provided unprecedented precision in astrometric properties of white dwarfs, such as their distances and proper motions, facilitating more accurate modelling of their evolutionary signatures \citep{Rowell19}. Recent studies, including those by \citet{Bergeron19} and \citet{Tremblay20}, have employed sophisticated atmosphere models to constrain the physical parameters of white dwarfs, providing new insights into their crystallisation process, magnetic fields, and the contribution of heavy elements to their atmospheric composition. Additionally, observational efforts such as those by \citet{Kilic10, Kilic17} have refined white dwarf ages and Galactic population types, offering crucial constraints on their cooling ages and formation histories.  These advances have significantly improved our understanding of white dwarfs as crucial tracers of stellar populations and as key players in Galactic evolution.

Estimating stellar distance is a crucial parameter for understanding the structure, formation and evolution of the Milky Way. This parameter provides insights into the chemical composition and dynamic processes that have shaped stellar populations over cosmic time. One of the most effective ways to derive stellar metallicity is through spectroscopic methods, which can yield detailed chemical abundances \citep[e.g.,][]{Karaali03b, Karaali11, TuncelGuctekin16, Celebi19}. However, spectroscopic analyses are limited to nearby stars because they rely on high-resolution spectra. Alternatively, photometric techniques, which use broad-band filters to measure stellar colours, allow for studying stars at much larger distances \citep{Bilir05, Bilir08, Bilir09}. Although photometric methods may offer lower precision than spectroscopy, they are invaluable for surveying distant stellar populations and contribute significantly to the understanding of the Galactic evolution \citep[c.f.,][]{Juric08, Ivezic08}.

For nearby stars, trigonometric parallax measurements provide highly reliable distance estimates. In particular, data obtained from the {\it Hipparcos} mission \citep{ESA97} play a crucial role in accurately determining stellar distances. Launched in 1989 by the European Space Agency (ESA), the {\it Hipparcos} satellite obtained high-precision parallax measurements for approximately 118,000 stars \citep{vanLeeuwen07}. However, due to their typically low luminosities, the sample of white dwarfs in the {\it Hipparcos} database remained limited \citep{Provencal98}. Despite this limitation, {\it Hipparcos} data provided parallax accuracy at the level of 10 micro arcsecond (mas) for some nearby white dwarfs, contributing significantly to the precise determination of their absolute magnitudes. 

The {\it Gaia} mission aims to construct the most comprehensive astrometric catalogue, encompassing the positions, proper-motion components, trigonometric parallaxes, and radial velocities of billions of stars in the Milky Way \citep{GaiaDR2}. Notably, the second \citep[{\it Gaia} DR2,][]{GaiaDR2} and third \citep[{\it Gaia} DR3,][]{GaiaDR3} data releases of {\it Gaia} have provided high-precision parallax measurements at the mas level for a vast sample of stars, including white dwarfs \citep{GentileFusillo19}. {\it Gaia} data have significantly refined the positioning of white dwarf stars in the Hertzsprung-Russell (HR) diagram, enabling more rigorous testing of theoretical models related to their evolutionary pathways and mass-luminosity relationships \citep{Tremblay24}. Furthermore, the high-accuracy parallaxes provided by {\it Gaia} have contributed to more precise calculations of the Galactic distribution and age estimations of the white dwarf population \citep{Bergeron19}.

Although space-based astrometric observations can be conducted with the {\it Gaia} satellite, the distances of faint stars cannot be accurately and precisely estimated due to biases in trigonometric parallax measurements as objects become fainter \citep{Luri18}. To overcome this problem, it is necessary to use relationships that are sensitive to the photometric colour indices of stars with accurately and precisely measured distances \citep[see also,][]{Bilir08, Bilir09}. Through these relationships, the photometric parallaxes of faint stars can be determined, allowing the estimation of their distances. Recent advancements in {\it Gaia} DR3 data and SDSS photometry have provided new opportunities to refine absolute magnitude determinations for a wide range of stellar populations, including white dwarfs. In this study, we utilise the {\it Gaia} DR3 catalogue \citep{GaiaDR3}, which provides precise distance and photometric data, to derive absolute magnitudes for white dwarfs. Additionally, we employ the SDSS colour index to create three distinct calibrations for absolute magnitudes, each based on different colour indices. These new calibrations offer improved precision over previous methods and enable the investigation of white dwarf populations across the Milky Way. By comparing these calibrations with values from the existing literature, we aim to refine our understanding of white dwarf properties and their role in the broader context of stellar evolution.
 
\section{Data}

The white dwarf stars analysed in this study were selected from the catalogue of \citet{Anguiano17}. This catalogue includes 20,247 white dwarfs with hydrogen-rich atmospheres, compiled from SDSS Data Release 12 \citep{Alam15}, along with their atmospheric model parameters, masses, ages, photometric distances, and radial velocities. Since the study by \citet{Anguiano17} was conducted in the early {\it Gaia} era, it does not incorporate data from the {\it Gaia} mission \citep{Gaia16mission}. Trigonometric parallax data for the white dwarfs in the catalogue were obtained by cross-matching with the {\it Gaia} DR3 catalogue \citep{GaiaDR3}, using a matching radius of $r\leq 5''$, which was adopted as the maximum separation between sources in the two catalogues. This resulted in a sample of 17,680 white dwarfs. To ensure data reliability, specific selection criteria were applied from the {\it Gaia} DR3 catalogue, including $\varpi > 0$, $\text{RUWE} < 1.4$, $\text{Dup} = 0$, and $\text{varflag} = \text{nonvariable}$. These criteria excluded potential binary and variable stars, refining the sample to 16,207 white dwarfs with measured trigonometric parallaxes. To establish a precise CMR for white dwarfs, an additional constraint on the relative parallax error as $\sigma_\varpi / \varpi \leq 0.10$ was applied, and the sample was reduced to 7,289 white dwarfs. 

\begin{figure*}
\centering
\includegraphics[width=\textwidth]{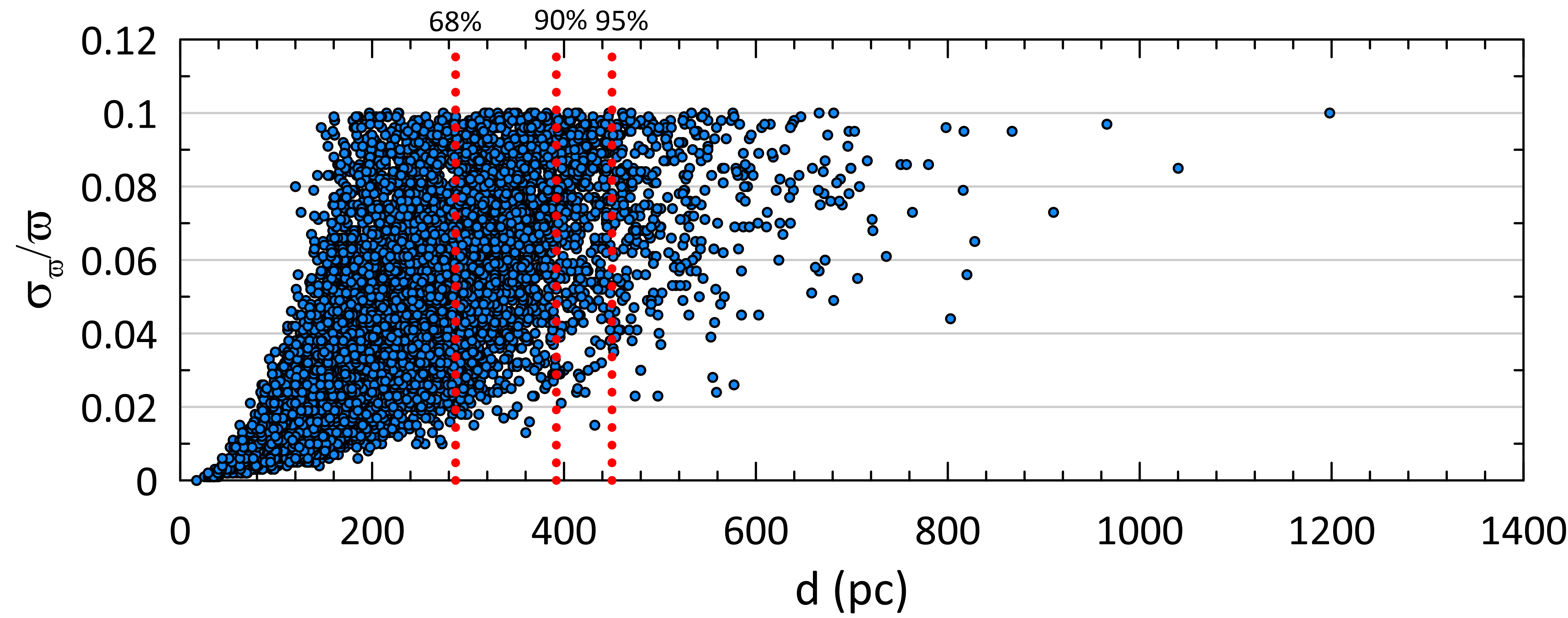}
\caption{Relative parallax errors ($\sigma_{\varpi}/\varpi$) as a function of distance for 7,289 white dwarf stars, based on trigonometric parallaxes from {\it Gaia} DR3. The red dotted lines indicate the distance thresholds within which 68\%, 90\%, and 95\% of the sample stars are located.}
\label{Fig:01}
\end {figure*}

The trigonometric parallax measurements of the selected white dwarf stars, obtained from the Gaia DR3 catalog, were converted into distances using the relation $d({\rm pc})=1000/\varpi$. Based on these data, a diagram was constructed according to the relative parallax errors of the stars. As shown in Figure~\ref{Fig:01}, the WD stars are distributed up to a distance of approximately 1250 pc. Moreover, it is observed that the estimated distances increase with increasing relative parallax errors. Additionally, it was determined that 68\%, 90\%, and 95\% of the white dwarf stars in the sample lie within distances of 287, 392, and 450 pc, respectively.

\section{Analysis}
\subsection{Photometric Absorptions Determination}
\begin{figure*}
\centering
\includegraphics[scale=0.50, angle=0]{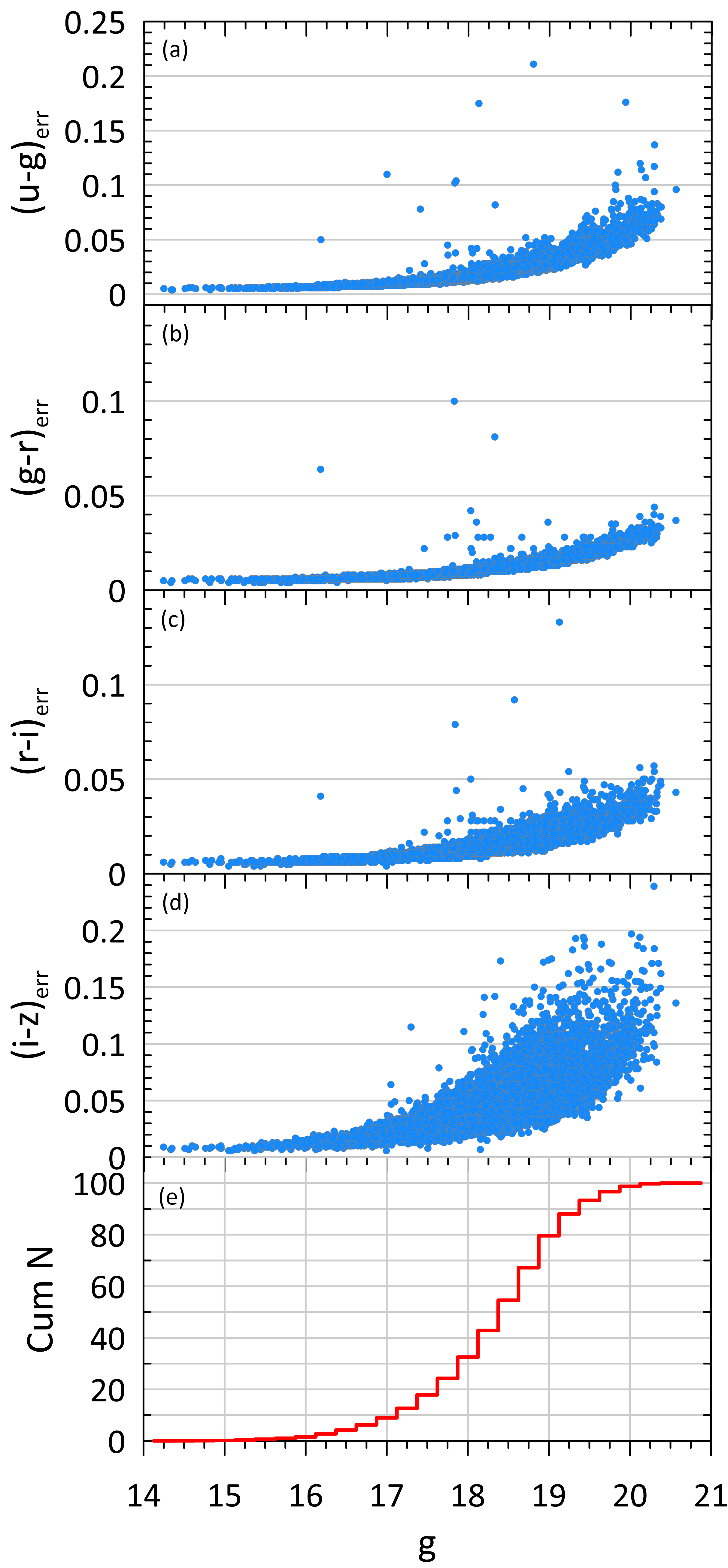}
\caption{The variation of the colour index errors of selected white dwarf stars with $g$-apparent magnitudes. (a) $(u-g)_{\rm err}\times g$, (b) $(g-r)_{\rm err}\times g$, (c) $(r-i)_{\rm err}\times g$, (d) $(i-z)_{\rm err}\times g$, and (e) cumulative distribution of the star sample.}
\label{Fig:02}
\end {figure*}

In this study, the uncertainties of the selected white dwarf stars in three colour indices as a function of the $g$-apparent magnitude in the SDSS system are plotted in Figure 1 to show the accuracy of colour index measurements in the SDSS photometric system. Based on SDSS photometric data, the 7,289 white dwarf stars analysed in this study have $g$-apparent magnitudes in the $14.3<g<20.6$ interval, with measurement uncertainties increasing toward fainter magnitudes. As white dwarfs are most effectively observed at shorter optical wavelengths, their colour index uncertainties generally range between 0.05 and 0.1 mag. However, for the $i-z$ colour index, which is defined at longer optical wavelengths, uncertainties can reach up to 0.2 mag. The bottom panel of Figure~\ref{Fig:02} represents the cumulative distribution of the selected white dwarfs. The median apparent magnitude of the cumulative distribution is approximately 18.5 mag, with 10\% to 90\% of the white dwarfs in the sample falling within  $17<g~({\rm mag})<19.3$ interval.

Since the study utilizes the $ugriz$ filters of the SDSS photometric system, it is necessary to correct the observed white dwarfs for interstellar extinction. For this purpose, the Galactic dust map of \citet{Schlafly11} was used. The $V$-band extinction values along the line of sight for the remaining 7,289 white dwarf stars from the main catalogue were obtained from the NASA IPAC database\footnote{\protect\url{https://irsa.ipac.caltech.edu/applications/DUST/}}. Since the $V$-band extinction values in the Galactic dust map are valid up to the boundary of the Milky Way, the extinction must be scaled according to the distance between the star and the Sun. The reduced $V$-band extinction was determined using the relation of \cite{Bahcall80}. 

\begin{equation}
A_{\rm d}(V)=A_{\infty}(V)\times \left[1-\exp\left(\frac{-|d\times\sin b|}{H}\right)\right],
\label{equ: Eq1}
\end{equation} 
here, $b$ (in degree) represents the Galactic latitude of the star, and $d$ (in pc) denotes its distance, determined using the trigonometric parallax measurements from the {\it Gaia} DR3 catalogue via the relation $d{\rm (pc)}=1000/\varpi$ (mas). The parameter $H$ corresponds to the dust scale height \citep[$H=125^{+17}_{-7}$ pc,][]{Marshall06}, while $A_{\infty}(V)$ represents the $V$-band extinction integrated along the line of sight to the Galactic boundary. The extinction value corresponding to the distance between the Sun and the star is given by $A_{\rm d}(V)$. 

The photometric bands in the SDSS  and {\it Gaia} DR3 catalogue were corrected for interstellar extinction using the selective absorption coefficients provided by \citet{Cardelli89} and \citet{O'Donnell94}. In this analysis, the extinction coefficients of these filters were derived based on the $R_V=3.1$ extinction curve from \cite{Cardelli89}. The central wavelengths of the SDSS passbands for $u$, $g$, $r$, $i$, and $z$ are 3561.79 \AA, 4718.87 \AA, 6185.19 \AA, 7499.70 \AA, and 8961.49 \AA, respectively, with the corresponding extinction coefficients $A_{\lambda}/A_{\rm V}$ calculated as 1.57465, 1.22651, 0.86639, 0.68311, and 0.48245. Likewise, the central wavelengths for the {\it Gaia} passbands corresponding to $G$, $G_{\rm BP}$, and $G_{\rm RP}$ are 6390.21 \AA, 5182.58 \AA, and 7825.05 \AA, respectively, with extinction coefficients $A_{\lambda}/A_{\rm V}$ determined as 0.83627, 1.08337, and 0.63439 \citep[see also,][]{Iyisan25}. Consequently, the following equations were applied to correct for interstellar absorption in the five SDSS and three {\it Gaia} passbands:

\begin{eqnarray}
u_{\rm 0} = u-A_{\rm u} = u - 1.57465\times\ A_{\rm d}(V)\\ \nonumber
g_{\rm 0} = g-A_{\rm g} = g - 1.22651\times\ A_{\rm d}(V) \\ \nonumber
r_{\rm 0} = r-A_{\rm r} = r - 0.86639\times\ A_{\rm d}(V) \\ \nonumber
i_{\rm 0} = i-A_{\rm i} = i - 0.68311\times\ A_{\rm d}(V) \\ \nonumber
z_{\rm 0} = z-A_{\rm z} = z - 0.48245\times\ A_{\rm d}(V) \\ \nonumber
\label{equ: Eq2}
\end{eqnarray}
and
\begin{eqnarray}
G_{\rm 0} = G-A_{\rm G} = G - 0.83627\times A_{\rm d}(V)\\ \nonumber
(G_{\rm BP})_{0} = G_{\rm BP}-A_{\rm G_{\rm BP}} = G_{\rm BP} - 1.08337\times A_{\rm d}(V) \\ \nonumber
(G_{\rm RP})_{0} = G_{\rm RP}-A_{\rm G_{\rm RP}} = G_{\rm RP} - 0.63439\times A_{\rm d}(V) \\  \nonumber
\label{equ: Eq3}
\end{eqnarray}

\begin{figure*}
\centering
\includegraphics[scale=0.7, angle=0]{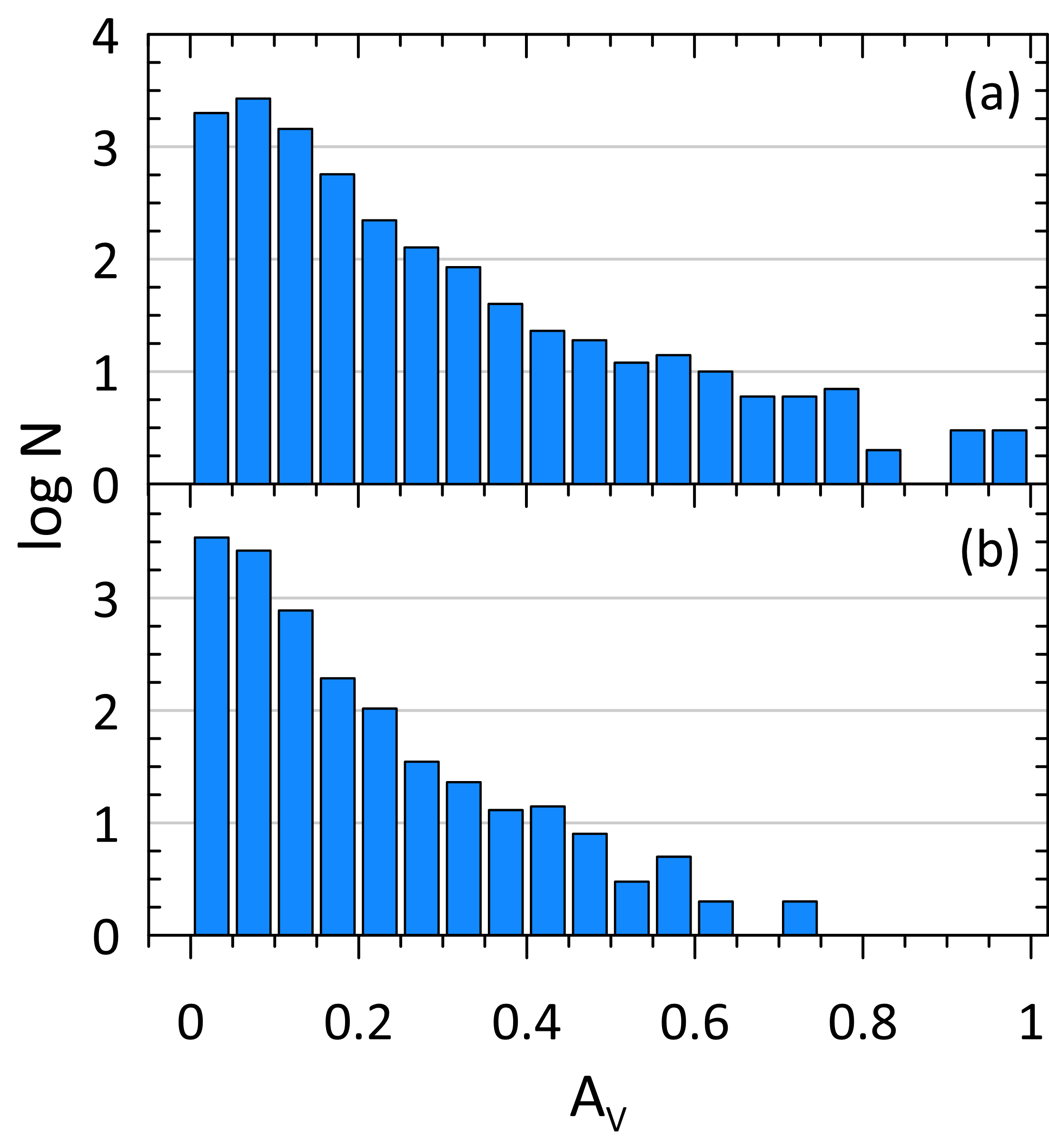}
\caption{Histograms of the original ($A_{\infty}(V)$) (a) and reduced absorption
($A_{{\rm d}}(V)$) values (b) of selected 5,516 white dwarf stars.} 
\label{Fig:03}
\end {figure*}

The $V$-band extinctions from the \citet{Schlafly11} dust maps for the selected white dwarf stars were shown in the upper panel of Figure~\ref{Fig:03}, and the estimated distance between the star and the Sun in $V$-band absorption were also represented in the lower panel of Figure~\ref{Fig:03}.

\subsection{HR Diagram of the White Dwarfs}
\begin{figure*}
\centering
\includegraphics[scale=0.80, angle=0]{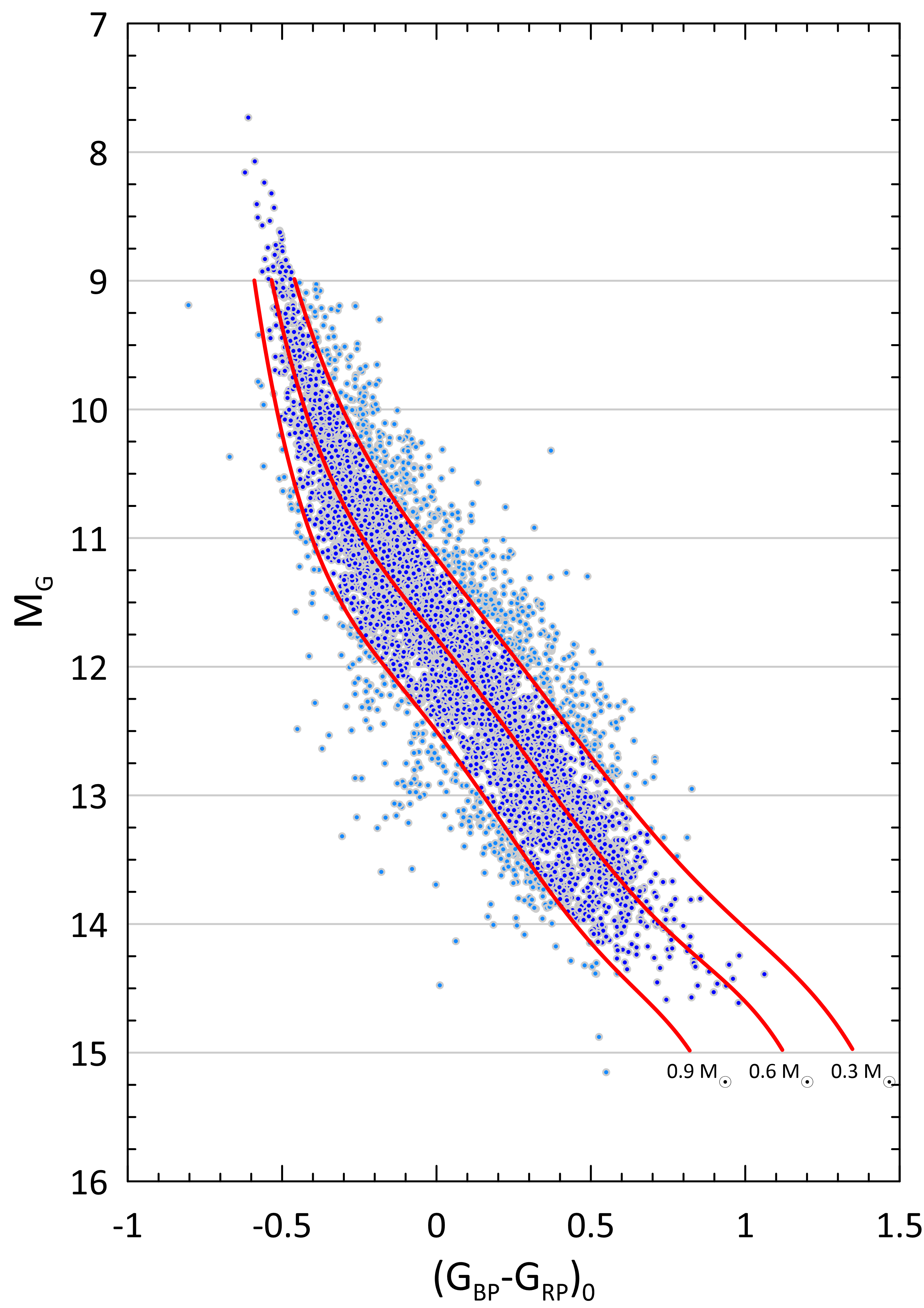}
\caption{The positions of the 5,516 white dwarf stars on the $M_{\rm G}\times (G_{\rm BP}-G_{\rm RP})$ CMD. The red curves show three different mass tracks from \citet{Holberg06}. Dark and light blue dots represent selected and scattered white dwarf stars.} 
\label{Fig:04}
\end {figure*}

In this study, the positions of selected white dwarfs on the colour-magnitude diagram (CMD) constructed from {\it Gaia} photometric and astrometric data were examined. The absolute magnitudes $M_G$ of the stars were estimated using the following equation

\begin{equation}
M_G = G_0 - 5 \times \log \left(\frac{1000}{\varpi} \right) + 5,
\label{equ: Eq4}
\end{equation}
where $G_0$ is extinction correction and  $\varpi$ is the trigonometric parallax. In the CMD shown in Figure~\ref{Fig:04}, it was determined that fall within the range $-0.80 < (G_{\rm BP} - G_{\rm RP})~{\rm(mag)}< 1.1$, and their absolute magnitude interval $7.5 < M_{\rm G}~{\rm(mag)} < 15.5$. Although the white dwarfs in the constructed CMD appear to be concentrated along the white dwarf sequence, they exhibit some scattering on the diagram. Given the high precision of the photometric and astrometric data of the white dwarf sample selected from the {\it Gaia} DR3 catalogue, this scattering may be attributed to unresolved binary systems. To minimise this effect, theoretical tracks of DA white dwarfs with masses of $0.3$, $0.6$, and $0.9 \ M_\odot$, as proposed by \citet{Holberg06}, were calibrated on the CMD. White dwarfs that fall outside these mass tracks were excluded from the statistical analysis of the colour-magnitude study. After applying the final constraints, 1,773 white dwarf stars were excluded from the analysis, and 5,516 white dwarfs were used in the calibration calculations.

\begin{figure*}
\centering
\includegraphics[scale=0.30, angle=0]{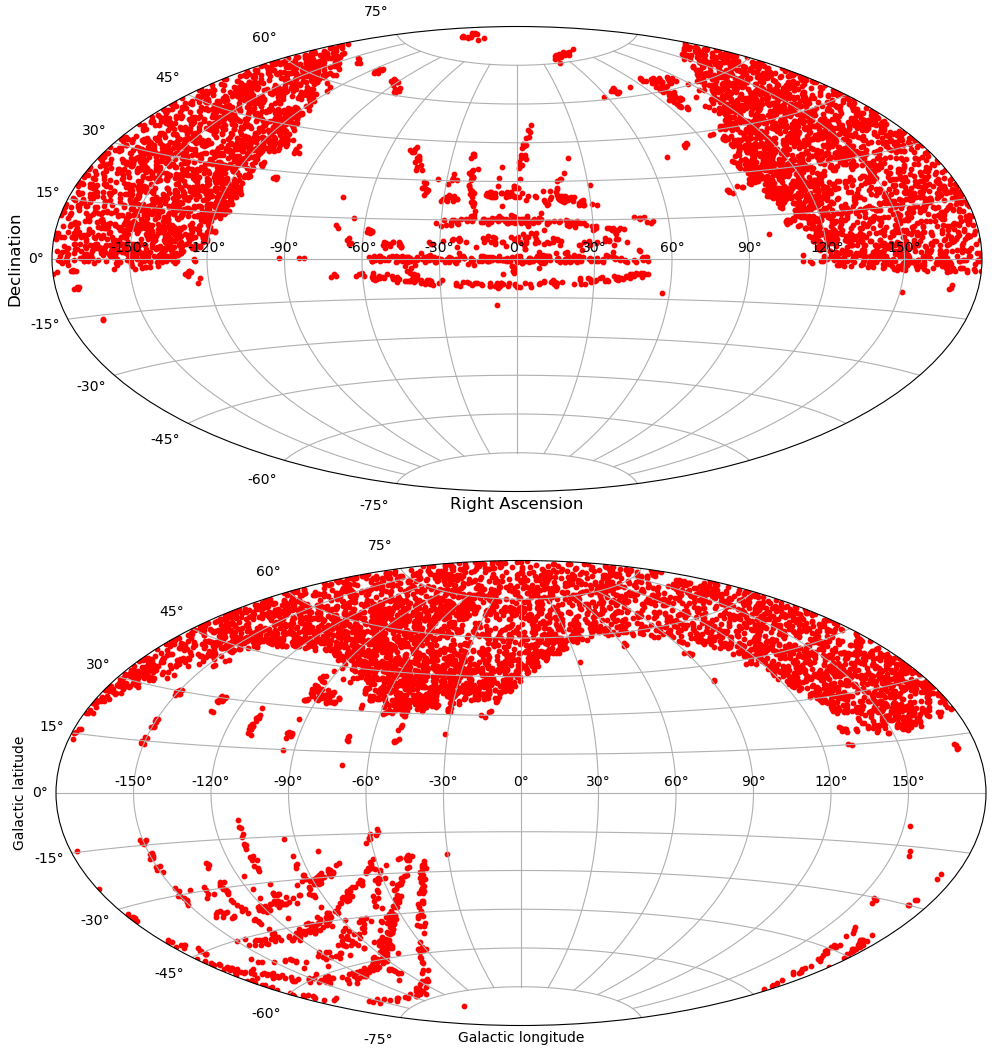}
\caption{The locations of 5,516 white dwarf stars in equatorial (top panel) and Galactic (bottom panel) coordinates.} 
\label{Fig:05}
\end {figure*}

To analyse the spatial distributions of the selected white dwarf sample, the stars were placed in both the equatorial and Galactic coordinate systems (Figure~\ref{Fig:05}). Since SDSS observations are carried out in Apache Point Observatory (APO)\footnote{\url{https://www.apo.nmsu.edu/}}, most of the white dwarfs are located in the northern hemisphere. However, due to the about $63^{\rm o}$ between the equatorial and Galactic coordinate systems, it was determined that approximately 14.3\% of the selected white dwarfs are located in the Galactic southern hemisphere (see lower panel of Figure~\ref{Fig:05}). Taking into account the trigonometric parallaxes of the 5,516 selected white dwarfs taken from the {\it Gaia} DR3 catalogue, their distances were calculated using the relation $d {\rm (pc)} = 1000/\varpi$ (mas). Based on these calculations, the distances of the white dwarfs relative to the Sun were found to be within the range $17 \leq d~{\rm (pc)} \leq 793$, with a median distance of $d_{\varpi} = 219$ pc.

\subsection{Colour-Magnitude Relations for SDSS White Dwarfs}
\begin{figure*}
\centering
\includegraphics[scale=0.45, angle=0]{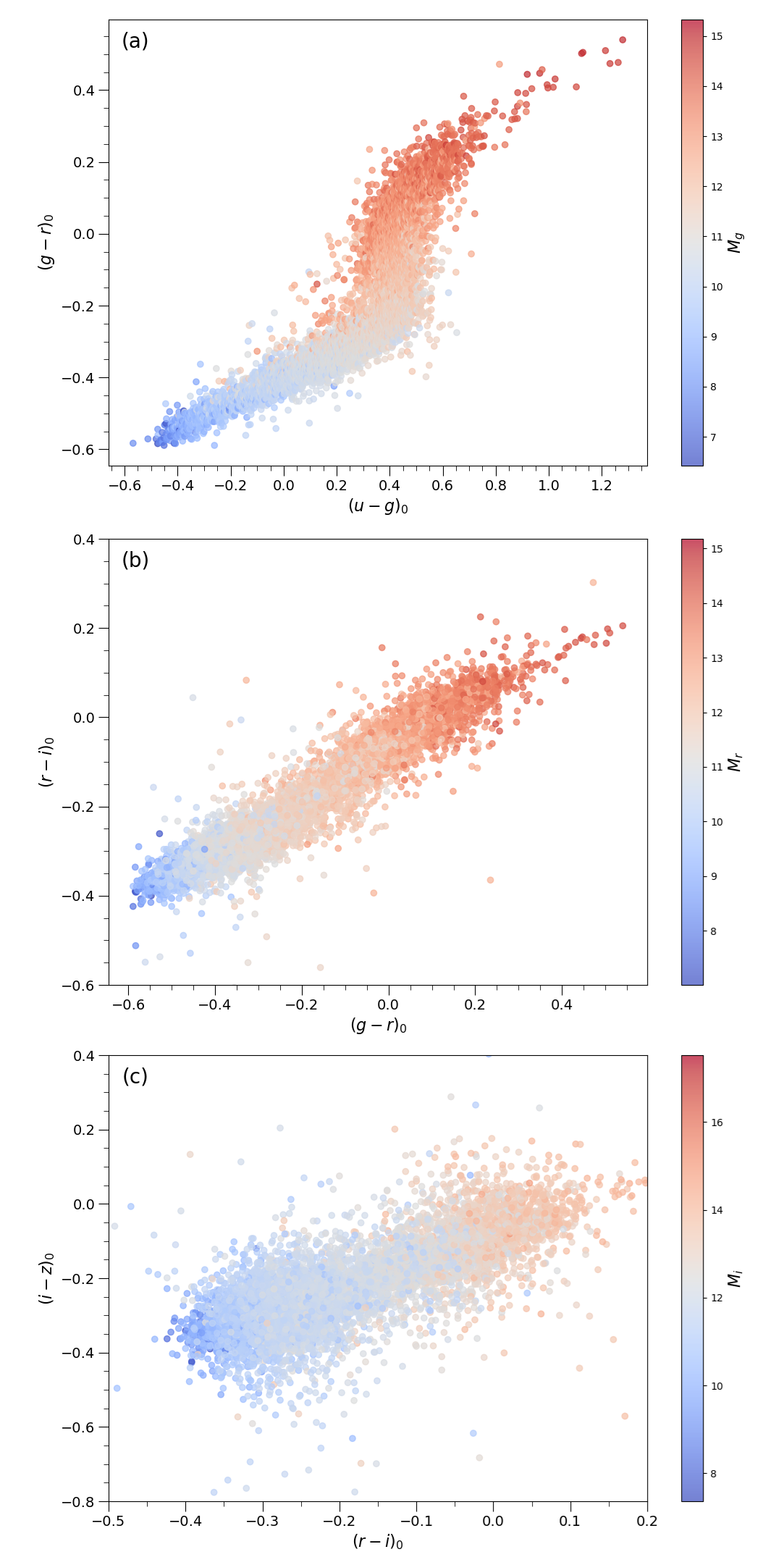}
\caption{The positions of the 5,516 white dwarf stars on the two-colour diagrams as a function of their absolute magnitudes. (a) $(u-g)_0 \times (g-r)_0$, (b) $(g-r)_0 \times (r-i)_0$, and (c) $(r-i)_0 \times (i-z)_0$.} 
\label{Fig:06}
\end {figure*}

This study aims to determine the  (CMRs) for white dwarf stars observed with SDSS photometry and with trigonometric parallax data available in the {\it Gaia} DR3 catalogue \citep{GaiaDR3}. In order to establish such relationships, it is necessary to examine the connection between the colour indices and absolute magnitudes of white dwarfs. In this context, the CMRs were analysed as a function of SDSS absolute magnitudes, which were derived from two-colour indices measured in the SDSS photometric system and trigonometric parallax data obtained from the {\it Gaia} DR3 catalogue. In this study, the positions of the selected white dwarf stars in three different two-colour diagrams (TCDs) are presented as functions of their absolute magnitudes in Figure~\ref{Fig:06}. As shown in the TCDs in Figure~\ref{Fig:06}, the absolute magnitudes become fainter as the colour index shifts from blue to red among white dwarfs. This finding provides strong evidence that the colour indices of the selected white dwarf stars are correlated with their absolute magnitudes. In this study, the following equations were used to establish the relationships between the colour indices and the absolute magnitudes of white dwarf stars.

\begin{equation}
M_{\rm g} = a_1 (u-g)_0^2 + b_1 (g-r)_0^2 + c_1 (u-g)_0 (g-r)_0 + d_1 (u-g)_0 + e_1 (g-r)_0 + f_1
\label{equ:Eq5}
\end{equation}

\begin{equation}
M_{\rm r} = a_2 (g-r)_0^2 + b_2 (r-i)_0^2 + c_2 (g-r)_0 (r-i)_0 + d_2 (g-r)_0 + e_2 (i-r)_0 + f_2
\label{equ:Eq6}
\end{equation}

\begin{equation}
M_{\rm i} = a_3 (r-i)_0^2 + b_3 (i-z)_0^2 + c_3 (r-i)_0 (i-z)_0 + d_3 (r-i)_0 + e_3 (i-z)_0 + f_3
\label{equ:Eq7}
\end{equation}
Similar relationships have been used in determining the absolute magnitudes of main-sequence stars in the {\it BVRI} and 2MASS ({$JHK_{\rm s}$}) photometric systems by \citet{Bilir08}, in the SDSS ({\it ugriz}) photometric system by \citet{Bilir09}, and in the {\it UBV} photometric system by \citet{Celebi19}.

In the sample consisting of 5,516 white dwarfs, the absolute magnitudes of the stars were analysed using multiple regression analyses, considering the variables associated with two colour indices in the SDSS photometric system. As a result of the statistical analyses, the parameters of the variables were estimated along with their uncertainties, which are presented in Table ~\ref{tab: Table1}. Statistical analyses indicate that the parameters calculated in the regression analyses are highly precise when their associated errors are taken into account. The fact that the probabilities ($P$) corresponding to the $T$-score values of the parameters are zero, and that the coefficient of determination ($R^2$) and standard deviation ($\sigma$) values calculated for the absolute magnitudes $M_{\rm g}$, $M_{\rm r}$, and $M_{\rm i}$ are determined as 
(0.951, 0.263), (0.943, 0.301), and (0.864, 0.368), respectively, suggest that the CMRs can be reliably and accurately utilised for distance determination.

\begin{table*}
\setlength{\tabcolsep}{4pt}
\centering
\caption{Coefficients and their errors for each equation and their coefficient of determination ($R^2$) and standard deviations ($\sigma$). In addition, the standard error, $T$, and $P$ values of each calculated parameter are given in the table.}
\label{tab:origin}
\begin{tabular}{lcccccccccccc} 
\hline
\hline
	& $a_1$	& $b_1$	& $c_1$	& $d_1$	& $e_1$	& $f_1$ & $R^2$ & $\sigma$ \\
\hline
$M_{\rm g}$ & -2.3287 & 1.0313 & 0.6356 & 1.5255   & 5.5578 & 12.4927 & 0.951 & 0.263 \\
Error       & 0.2381  & 0.3669 & 0.6020 & 0.2628   & 0.3094 & 0.0715  & ---   & ---   \\
$T$         & -9.78   & 2.81   & 1.06   & 5.80     & 17.96  & 174.730 & ---   & ---   \\
$P$         & 0.000   & 0.005  & 0.003  & 0.000    & 0.000  & 0.000   & ---   & ---   \\ 
\hline
            & $a_2$   & $b_2$  & $c_2$  & $d_2$    & $e_2$  & $f_2$   & $R^2$ & $\sigma$\\ 
\hline
$M_{\rm r}$	& -4.6110 & 0.25683 & 3.0200 & 3.36816 & 2.1026 & 13.0225 & 0.943 & 0.301 \\
Error       & 0.2318  & 0.0202  & 0.3790 & 0.0697  & 0.1234 & 0.0097  &  ---  & ---   \\
$T$         & -19.89  & 12.75   & 7.97   & 48.31   & 17.05  & 1342.24 &  ---  & ---   \\
$P$         & 0.000   & 0.000   & 0.000  & 0.000   & 0.000  &  0.000  &  ---  & ---   \\ 
\hline
	      & $a_3$	& $b_3$	& $c_3$	& $d_3$	& $e_3$	& $f_3$  & $R^2$ & $\sigma$\\ 
\hline
$M_{\rm i}$ & -2.4688 & 0.1963 & -5.3970 & 4.2925 & 0.3183 & 13.2530 & 0.864 & 0.368 \\	
Error       & 0.1704  & 0.0336 & 0.1912  & 0.1086 & 0.0822 & 0.0108  &  ---  & ---   \\
$T$         & -14.49  &  5.83  & -28.22  & 39.51  &  3.87  & 1226.35 &  ---  & ---   \\
$P$         & 0.000   & 0.000  & 0.000   & 0.000  & 0.000  &  0.000  &  ---  & ---   \\

\hline
\hline
\end{tabular}
\label{tab: Table1}
\end{table*}

\begin{figure*}
\centering
\includegraphics[scale=0.60, angle=0]{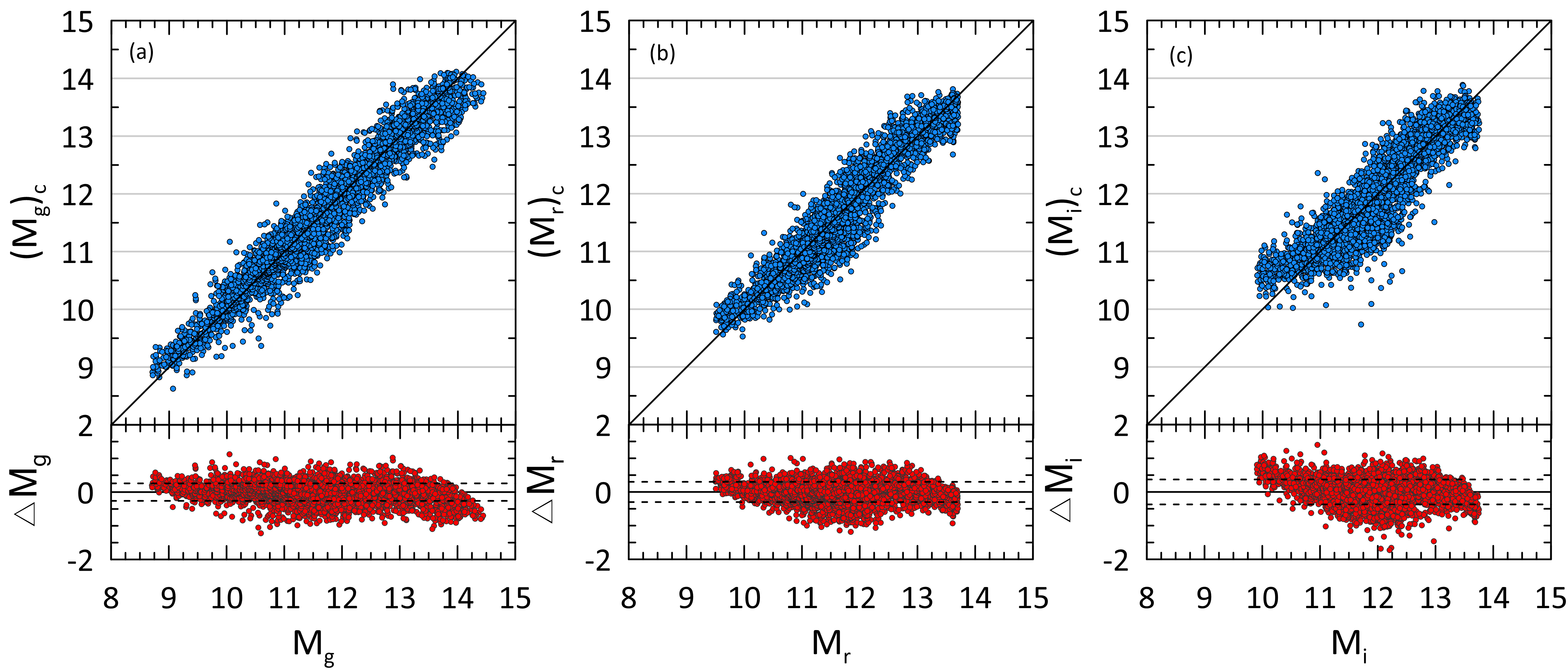}
\caption{Comparison of the calculated and the original SDSS absolute magnitudes (upper panels) and distribution of the absolute magnitude residuals ($\Delta M$) concerning the original absolute magnitudes (lower panels) for 5,516 white dwarf stars. The solid black line represents one-to-one lines, and the dashed lines show $\pm 1\sigma$ prediction levels.} 
\label{Fig:07}
\end {figure*}

The comparison between the CMRs obtained using the white dwarfs in the sample and those given by Equations~(\ref{equ:Eq5}), ~(\ref{equ:Eq6}), and (\ref{equ:Eq7}) with the original absolute magnitudes is presented in the upper panel of Figure~\ref{Fig:07}. The variation of the differences between the calculated and original absolute magnitudes as a function of the original absolute magnitudes is shown in the lower panel of Figure~\ref{Fig:07}. As seen in the figure, Equation~(\ref{equ:Eq5}), derived from the $(u-g)_{\rm 0}$ and $(g-r)_{\rm 0}$ colour indices, spans a wider absolute magnitude range compared to the CMRs of Equations~(\ref{equ:Eq6}) and (\ref{equ:Eq7}), which are obtained from the $(M_{\rm r}, M_{\rm i})$ colour indices. Among these three relations, the least scatter in the CMRs is observed when using the $ugr$ filters, where white dwarfs are most prominently detected. Including SDSS filters corresponding to longer wavelengths in the relation results in greater deviations of the calculated absolute magnitudes from the original values. This discrepancy arises from the fact that white dwarfs do not exhibit efficient luminosity at redder colours.

\section{Discussion}

In this study, the trigonometric parallaxes from the {\it Gaia} DR3 catalogue, the corrected distances obtained using the Bayesian method by \citet[BJ21;][]{BJ21}, and the distances provided in \citet[A17;][]{Anguiano17}, which serves as the primary data source of this study, were used to compare the CMRs derived for white dwarfs observed in SDSS filters with those available in the literature. To achieve this, three absolute magnitude values calculated in this study were used to determine the distances of white dwarfs from the Sun based on their apparent magnitudes via the distance modulus. While the distances of stars were determined from the trigonometric parallaxes provided in the {\it Gaia} DR3 catalogue using the relation $d {\rm (pc)} = 1000/\varpi$ (mas), the distances from BJ21 were obtained by extracting the corresponding values from the BJ21 catalogue based on the equatorial coordinates of the stars.  

\begin{figure*}
\centering
\includegraphics[scale=0.25, angle=0]{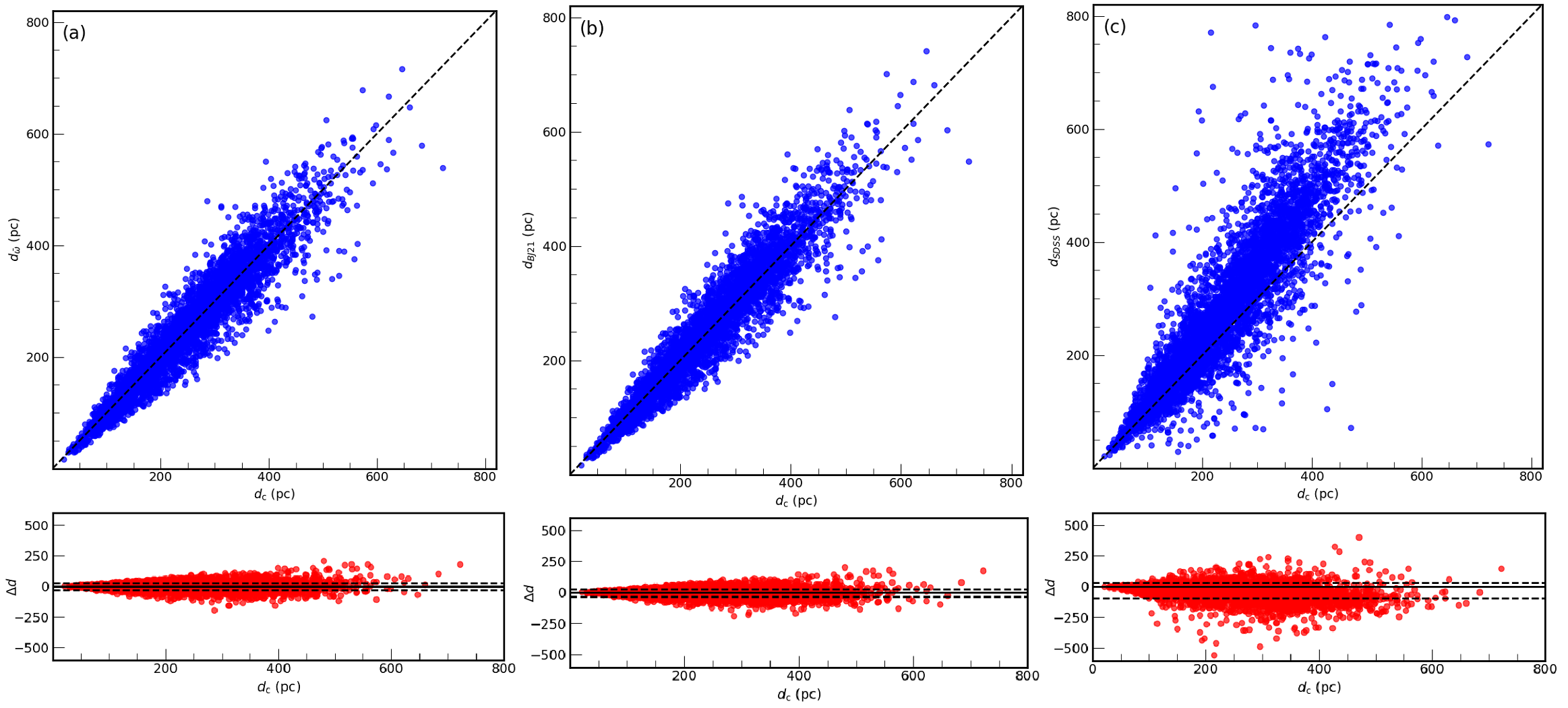}
\caption{Comparison of the distances to the 5,516 white dwarfs calculated in this study with those of Gaia (a), BJ21 (b), and A17 (c), respectively (top panel) and illustration of the distance differences (bottom panel). The solid black line represents one-to-one lines, and the dashed lines show $\pm 1\sigma$ prediction levels.} 
\label{Fig:08}
\end {figure*}

The comparison of the three distance datasets obtained from the literature with those calculated using the CMRs determined in this study is presented in Figure~\ref{Fig:08}. In the horizontal axes of the plots, the distances calculated in this study are shown, while the vertical axes represent the distances from {\it Gaia} DR3 (a), BJ21 (b), and A17 (c), respectively. In the lower panel of the figure, the horizontal axes again show the distances obtained in this study, whereas the vertical axis presents the differences between these distances and those from the literature. The median and standard deviations of the distance differences between the values obtained in this study and those from {\it Gaia} DR3, BJ21, and A17 were calculated as $(-1, 30)$, $(-2.4, 31)$, and $(-30, 63)$ pc, respectively. As seen in Figure~\ref{Fig:08}, the distances obtained using the CMRs in this study are highly consistent with those from {\it Gaia} DR3 and BJ21. However, a significant discrepancy is observed compared to A17, particularly in terms of the zero-point offset and standard deviation of the distances. The consistency of the distances calculated in this study with those provided by {\it Gaia} DR3 and BJ21 can be attributed to the high precision of the astrometric data used for the calibration stars. However, beyond 400 pc, a systematic bias is observed in the agreement between the distances obtained in this study and those from A17. It has been determined that A17 systematically overestimates the distances to white dwarfs.

\section{Summary \& Conclusion}

In this study, we examined the CMRs for DA-type white dwarfs observed in the SDSS photometric system. A sample of 20,247 white dwarfs from the \citet{Anguiano17} catalogue was cross-matched with {\it Gaia} DR3 \citep{GaiaDR3} to obtain trigonometric parallax measurements. The SDSS magnitudes of these stars were corrected for interstellar extinction using Galactic dust maps, and distances were estimated accordingly. To ensure high-precision calibrations, a subsample of 5,516 white dwarfs was selected based on brightness limits ($ g_0 < 21 $ mag) and relative parallax errors ($ \sigma_{\varpi} / \varpi < 0.1 $).  

Using this refined sample, three distinct CMRs were derived for absolute magnitudes $ M_{\rm g}$, $ M_{\rm r}$, and $ M_{\rm i}$, each based on two-colour indices. The coefficient of determinations of these relations ranged from 0.86 to 0.95, indicating a strong statistical reliability. Additionally, the standard deviations between the absolute magnitudes obtained from these relations and the original values varied from 0.26 to 0.37 mag. The analysis revealed that the relation constructed from the $(u-g)_{\rm 0}$ and $ (g-r)_{\rm 0}$ colour indices spans a broader absolute magnitude range than those based on redder filters, such as $(M_{\rm r}, M_{\rm i})$, which exhibit larger deviations due to the diminished luminosity of white dwarfs at longer wavelengths. A comparison of our computed distances with those provided in {\it Gaia} DR3, \citep{BJ21}, and \citet{Anguiano17} revealed a strong agreement with {\it Gaia} DR3 and \citep{BJ21}. The median differences between our derived distances and those from {\it Gaia} DR3 and \citep{BJ21} were found to be $-1$ pc and $-2.4$ pc, respectively, with standard deviations of 30 pc and 31 pc. However, a significant systematic offset was observed when comparing our results with \citet{Anguiano17}, particularly for stars beyond 400 pc, where \citet{Anguiano17} systematically overestimates distances. It is important to note that the {\it Gaia} DR3 \citep{GaiaDR3} and \citet{BJ21} distances are not fully independent, as {\it Gaia} data were also utilized in constructing the CMR employed in this study. In contrast, the distances from \citet{Anguiano17} are independent of the {\it Gaia}-based relations, thus providing an external benchmark for assessing the consistency and reliability of our results. This distinction is essential and should be considered when interpreting the comparative distance analyses.  

In this study, we have demonstrated that the CMRs calibrated for SDSS filters can be effectively utilised to determine the distances of stars that appear fainter in {\it Gaia} photometry. While the {\it Gaia} DR3 catalogue provides trigonometric parallax data for stars up to $G=21$ mag, the relatively large parallax errors significantly impact measurement accuracy and precision, thus affecting the derived distances of the analysed stars. Overall, this study presents improved calibrations for estimating the absolute magnitudes and distances of white dwarfs using SDSS photometry, offering valuable insights into Galactic structure and evolution. These new relations are valuable tools for refining white dwarf population studies and will facilitate future research in accurately determining distances for faint white dwarfs in large-scale astronomical surveys.

\section*{Acknowledgements}
We thank the anonymous referees for their insightful and constructive suggestions, which significantly improved the paper. This research has made use of NASA's Astrophysics Data System. This research has made use of the SIMBAD database, operated at CDS, Strasbourg, France. This work presents results from the European Space Agency (ESA) space mission Gaia. Gaia data are being processed by the Gaia Data Processing and Analysis Consortium (DPAC). Funding for the DPAC is provided by national institutions, in particular the institutions participating in the Gaia MultiLateral Agreement (MLA). The Gaia mission website is https://www.cosmos.esa.int/gaia. The Gaia archive website is https://archives.esac.esa.int/gaia. This research has made use of the NASA/IPAC Infrared Science Archive, which is funded by the National Aeronautics and Space Administration and operated by the California Institute of Technology.


\bibliographystyle{mnras}
\bibliography{WD-CMR.bib}

\bsp	
\label{lastpage}
\end{document}